\input{aipcheck}

\documentclass[
    ,final            
  ]
  {aipproc}

\layoutstyle{6x9}

\begin{document}

\title{Contribution of the HMPID detector to the high-p$_T$ physics at LHC}

\author{D. Di Bari}{
  address={Universit\'a degli Studi di Bari, Dipartimento Interateneo di Fisica "M. Merlin", Bari, Italy}
     ,altaddress={Istituto Nazionale di Fisica Nucleare, Sezione di Bari, Italy} 
}

\author{A. Mastroserio}{
    address={Universit\'a degli Studi di Bari, Dipartimento Interateneo di Fisica "M. Merlin", Bari, Italy}
     ,altaddress={Istituto Nazionale di Fisica Nucleare, Sezione di Bari, Italy} 
}

\author{L. Molnar }{
     address={Istituto Nazionale di Fisica Nucleare, Sezione di Bari, Italy} 
     ,altaddress={Res. Inst. Particle \& Nucl. Phys., Hungarian Academy of Sciences, Hungary \footnote{This work was partially supported by the grants OTKA
NK62044, IN71374.}} }

\author{E. Nappi}{
     address={Istituto Nazionale di Fisica Nucleare, Sezione di Bari, Italy} 
}

\author{D. Perrino}{
    address={Universit\'a degli Studi di Bari, Dipartimento Interateneo di Fisica "M. Merlin", Bari, Italy}
     ,altaddress={Istituto Nazionale di Fisica Nucleare, Sezione di Bari, Italy} 
}

\author{G. Volpe}{
     address={Istituto Nazionale di Fisica Nucleare, Sezione di Bari, Italy} 
}

\begin{abstract}
The LHC will deliver unexplored energy regimes for proton-proton and
heavy-ion collisions. As shown by the RHIC experiments, particle
identification over a large momentum range is essential to
disentangle physics processes, especially in the intermediate p$_T$
(1 $<p_{T}<5$ GeV/c) region. The novel design of the High-Momentum
Particle Identification Detector (HMPID), based on large surface CsI
photocathodes, is able to identify $\pi^{\pm}$, $K^{\pm}$, $p$ and
$\overline{p}$ in the momentum region where bulk medium properties
and hard scatterings interplay. Furthermore, measurement of
resonance particles such as the $\phi \rightarrow K^+K^-$ could
provide information on the system evolution. The HMPID layout and
segmentation are optimized to study particle correlations at high
momenta describing the early phase and the dynamical evolution of
the collision. At LHC, the increased hard cross section will
significantly be enhanced compared to RHIC. Jet reconstruction 
via Deterministic Annealing can address jet quenching and detailed measurements of jet
properties. In this paper, we present these selected topics from the
possible HMPID contributions to the physics goals of LHC.
\end{abstract}

\maketitle


\section{The High-Momentum Particle Identification Detector}

The High Momentum Particle Identification Detector (HMPID)
~\cite{tdr}, consists of seven identical Ring-Imaging Cherenkov
modules, covering 11 m$^2$ in total in a single arm arrangement. The
HMPID exploits the properties of Cherenkov light production and
large surface CsI photocathodes to identify charged particles on
track-by-track basis. Charged particles traversing the detector
produce Cherenkov photons in the C$_6$F$_{14}$ radiator
(<n>$\approx$1.292). The Cherenkov photons are detected by MWPCs.
One photocathode of the MWPC is the segmented CsI photocathode where
the typical ring pattern is formed. A dedicated algorithm of pattern
recognition gives information on the mean Cherenkov photon emission
angle ($\Theta_{C}$). Hence, the HMPID can identify charged pions
and kaons in the momentum range $p \sim$ 1 - 3 GeV/c and protons up
to $p~\sim$ 5 GeV/c with 3$\sigma$ separation. Detailed description
of the HMPID can be found in ~\cite{tdr}.

\begin{figure}
   \includegraphics[trim= 25mm 70mm 20mm 0mm, clip, scale=0.47]{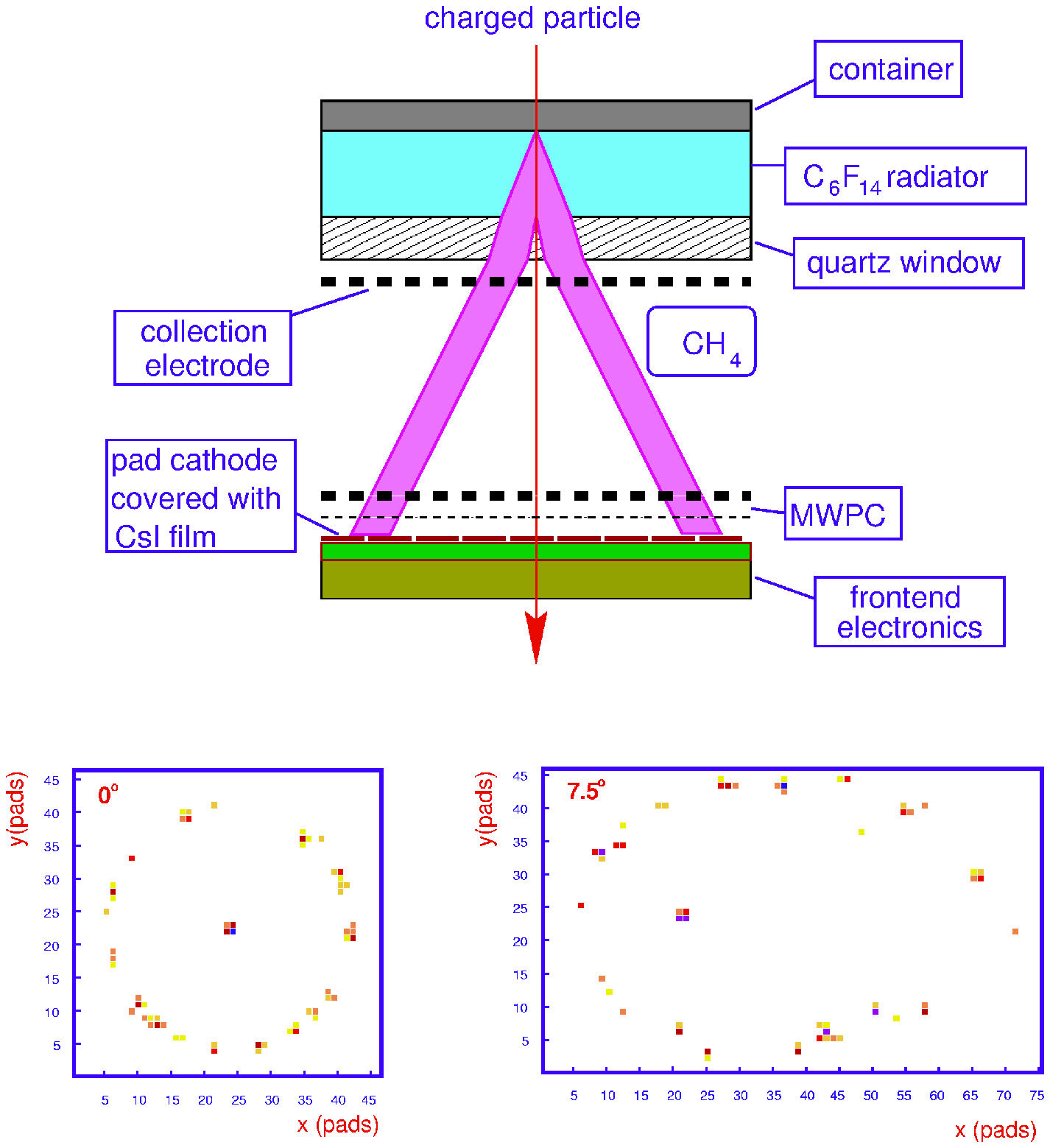}
   \includegraphics[scale=0.3]{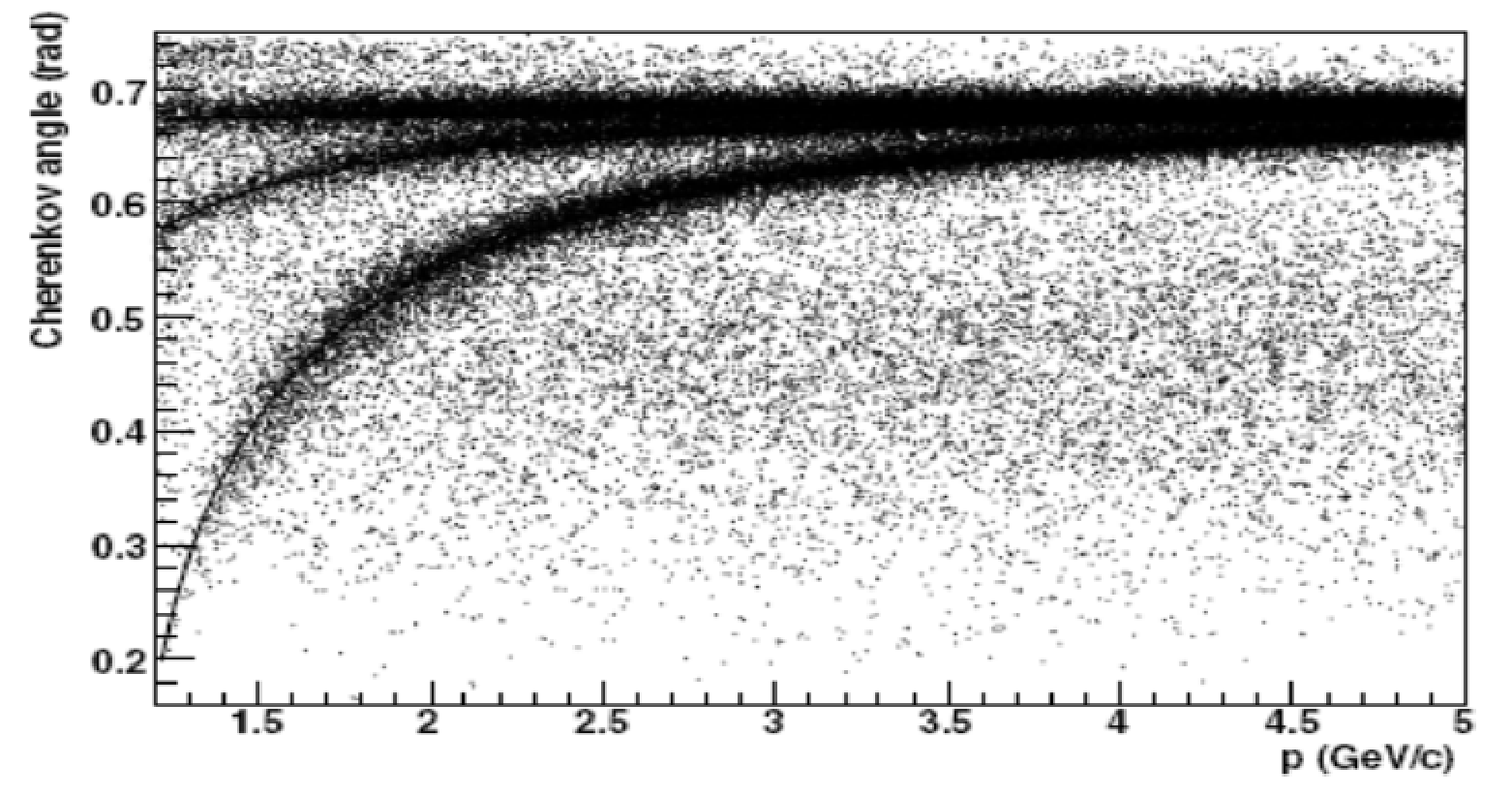}
   \caption{Left: Schematic view of the HMPID~\cite{tdr}.
   Right: Reconstructed Cherenkov angle in the HMPID as a function of the track
momentum. Equal concentration of charged pions, kaons and protons
have been merged with HIJING events (dN$_{\textrm{ch}}$/$\,$d$\eta$
= 6000). Solid lines indicate the predicted curves for $\pi$, $K$
and $p$~\cite{AlicePPR}.}
\end{figure}

\section{HMPID related observables: single hadron spectra and $\Phi$(1020)}

\begin{figure}
   \includegraphics[scale=0.47]{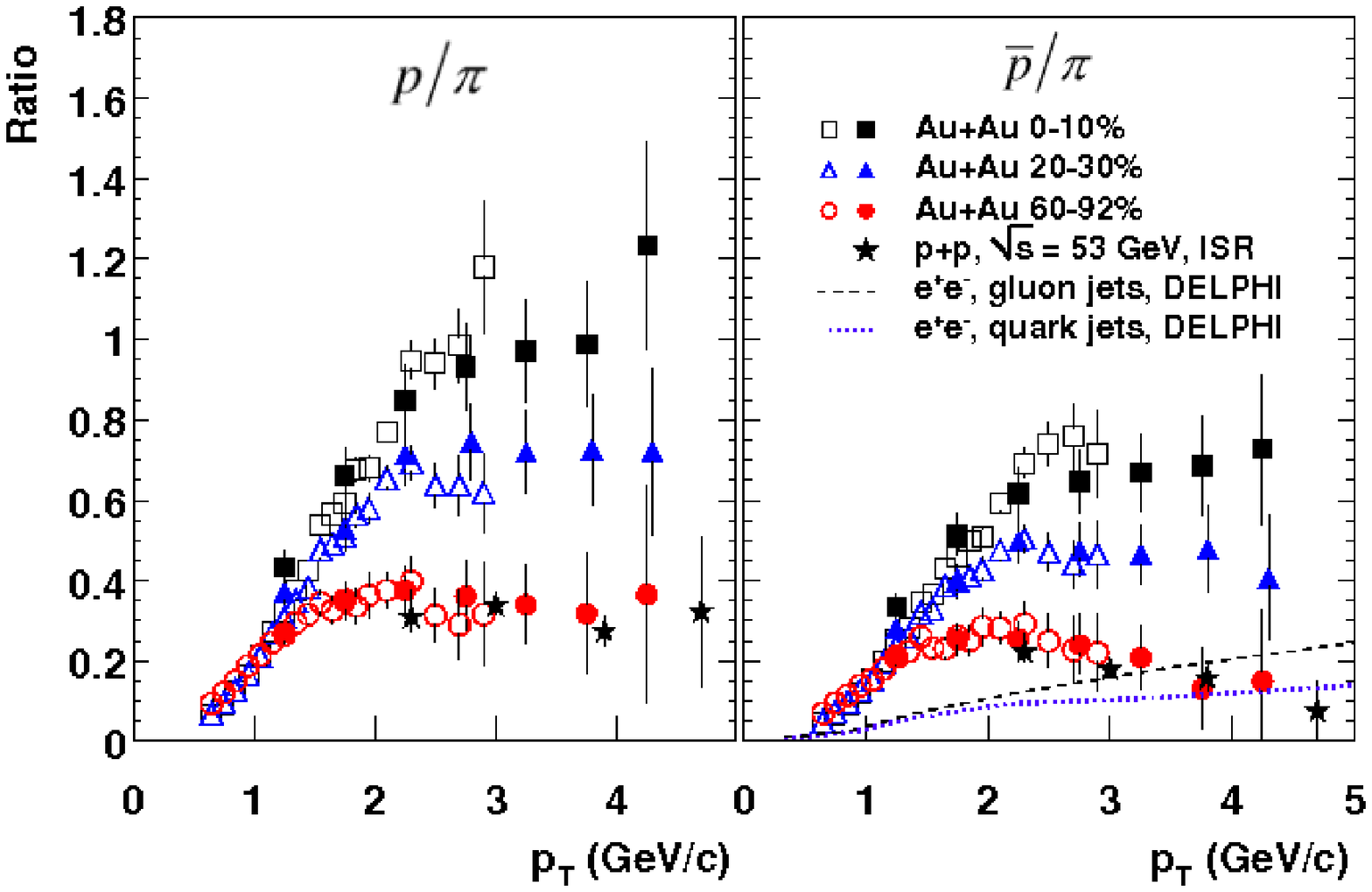}
   \hspace{-1cm}
   \includegraphics[scale=0.35]{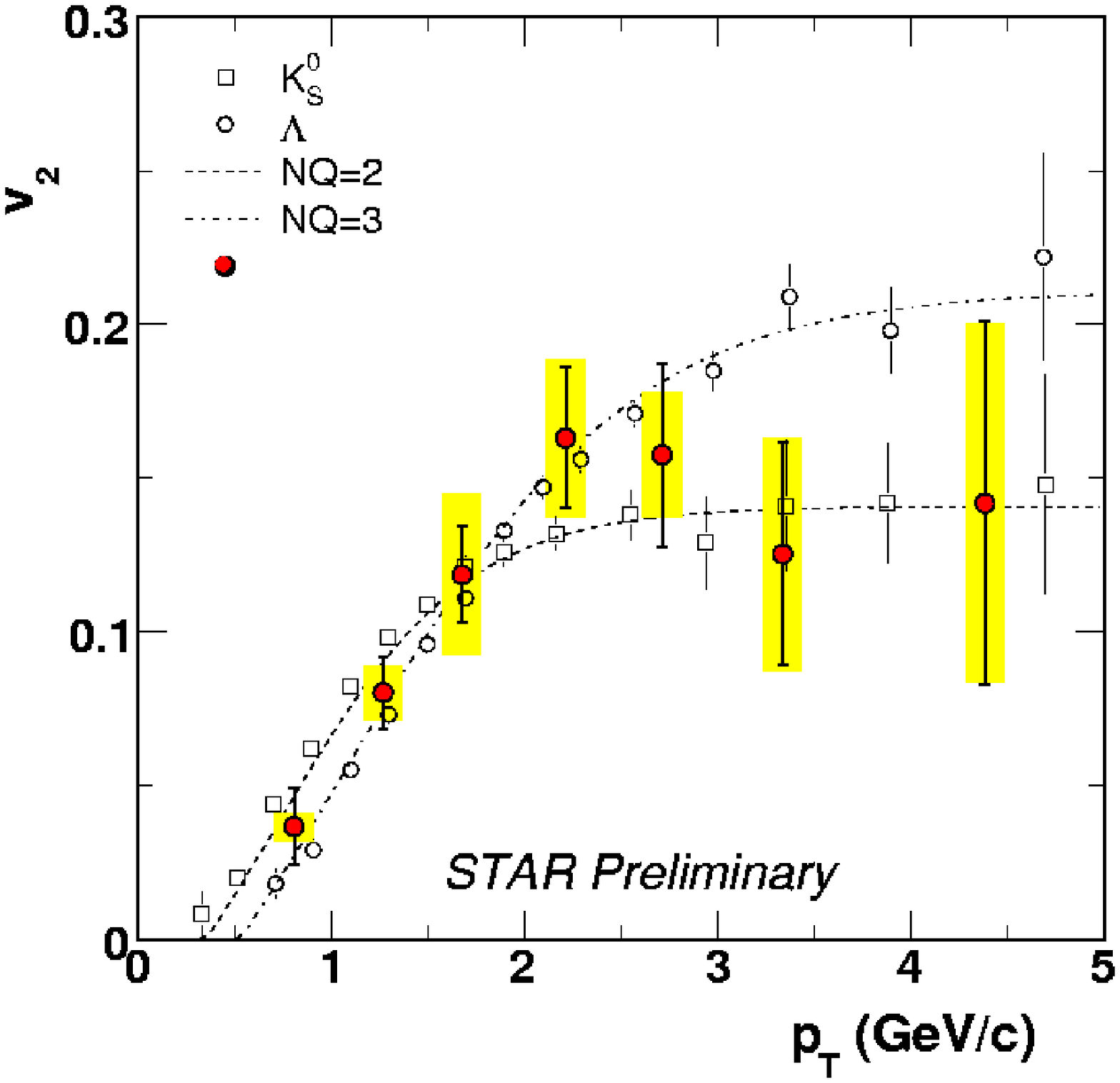}
   \caption{Left: The $p/\pi$ ratios are shown for different
   centralities in 200 GeV Au-Au collisions.
   The ratio reaches 1 in the most central collisions.
   A comparison with p-p collisions is also shown \cite{ppi}. Right: $\phi(1020)$
elliptic flow (filled circles) is shown with respect to the
$\Lambda$ and the $K^0$.
  The $v_2$ follows the baryon trend up to 2~GeV/$c$, then it behaves like the $K^0$
  \cite{v2}.}
   \label{fig1}
\end{figure}

\begin{figure}[b]
  \includegraphics[angle=90, scale=0.5]{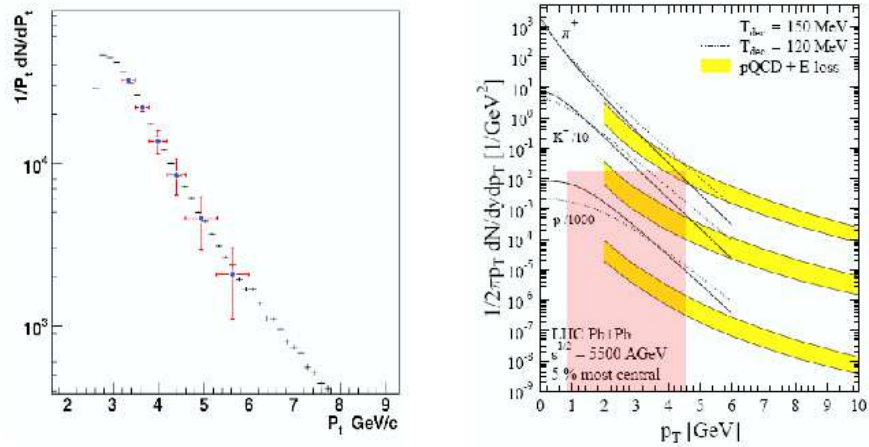}
  \caption{Left: Reconstructed momentum spectrum of the $\phi(1020)$ meson in its hadronic
  decay channel $\phi \rightarrow K^+ K^-$
  in central Pb-Pb collisions at 5.5 TeV by simulation. The kaons were identified by the HMPID between
  $p_T \approx$ 1.3 - 6 GeV/$c$. Right: Predictions for $\pi$, $K$ and $p$ momentum
  distributions at LHC energies \cite{pspec}.
  For protons there is no continuity between hydrodynamical curve and pQCD+energy loss results between 2 and 5
  GeV/$c$, the region where the HMPID can identify them.}
  \label{fig2}
\end{figure}

Some signals of the Quark Gluon Plasma (QGP) are influenced by the
mechanisms involved at the hadronization stage \cite{coal},
therefore it is important to disentangle such effects from the
deconfined state observables. The momentum region between 2 and 5
GeV/$c$ is crucial to this aim. At the Relativistic Heavy Ion
Collider (RHIC), where Au nuclei were collided at $\sqrt{s_{NN}}$ =
200 GeV, hadron production seems to be driven by quark content
dependent mechanisms. A striking result is that the baryon over
meson ratio reaches unity in most central collisions and the
mechanism of the quark coalescence at the hadronization describes
quite well such a behavior \cite{coal}. The left and the central
panel in Figure \ref{fig1} show the unexpected deviation of such a
ratio from p-p events where the deconfined state is not expected to
form. Furthermore, the elliptic flow of the expanding fireball
reaches different values depending on the quark content of the
hadrons (right panel in Figure \ref{fig1}). Within this scenario the
$\phi(1020)$ has a meson nature ($s \overline{s}$ bound state) and a
proton-like mass, therefore it is a good tool to probe the QGP and
the role of both mass values and quark content in its subsequent
hadronization phase. Strange quarks are mainly produced by gluon
fragmentation within the hot and dense medium, therefore the short
lifetime of $\phi$ and its small cross section with non strange
hadrons \cite{phiProbe} make this particle a good tool to probe both
the deconfined and the newly formed hot hadronic state. An
interesting result is the elliptic flow of $\phi$: it follows a mass
ordering, falling between the heavier $\Lambda$ and the $K^0$, up to
2 GeV/$c$ and then it follows the $K^{0}$ trend \cite{v2}. These
measurements suggest that a comparison between the $\phi(1020)$ and
the proton production can provide new insights on the effects of
both QGP and the first hot confined state and the HMPID detector can
identify $\phi$ mesons and protons right between 2 and 5 GeV/$c$
(Left plot in Figure \ref{fig2}).

Furthermore, recent theoretical results of hadron momentum
distribution at LHC energies show that between 2 and 5 GeV/$c$ the
predictions from hydrodynamical estimates for protons do not match
smoothly to the respective pQCD + partonic energy loss calculations,
within the uncertainty bands. This implies that future measurements
of the proton momentum distribution itself within this peculiar
momentum region will be very important to understand the dynamics of
heavy-ion collisions.

\begin{figure}
   \includegraphics[scale=0.4]{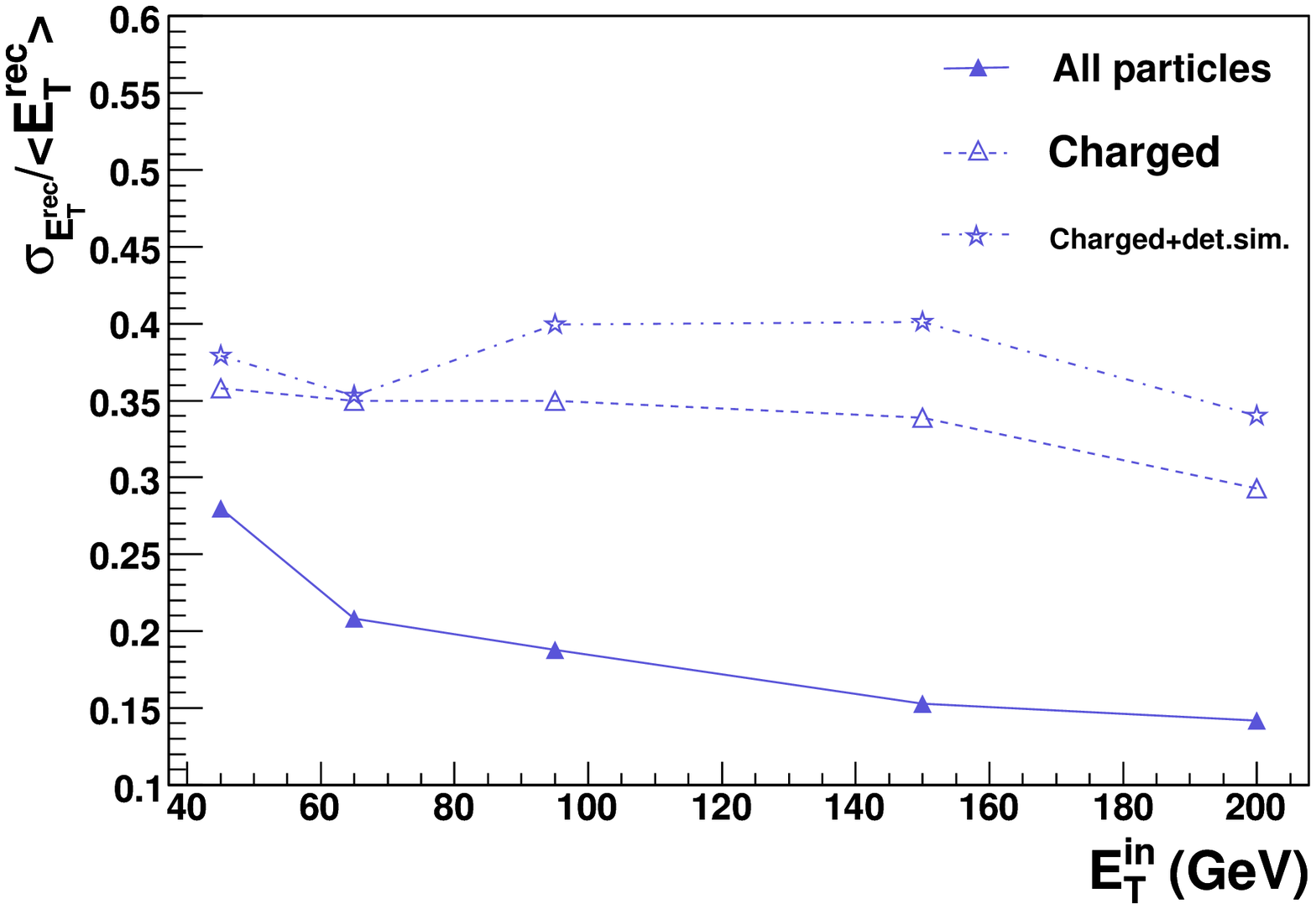}
   \includegraphics[scale=0.4]{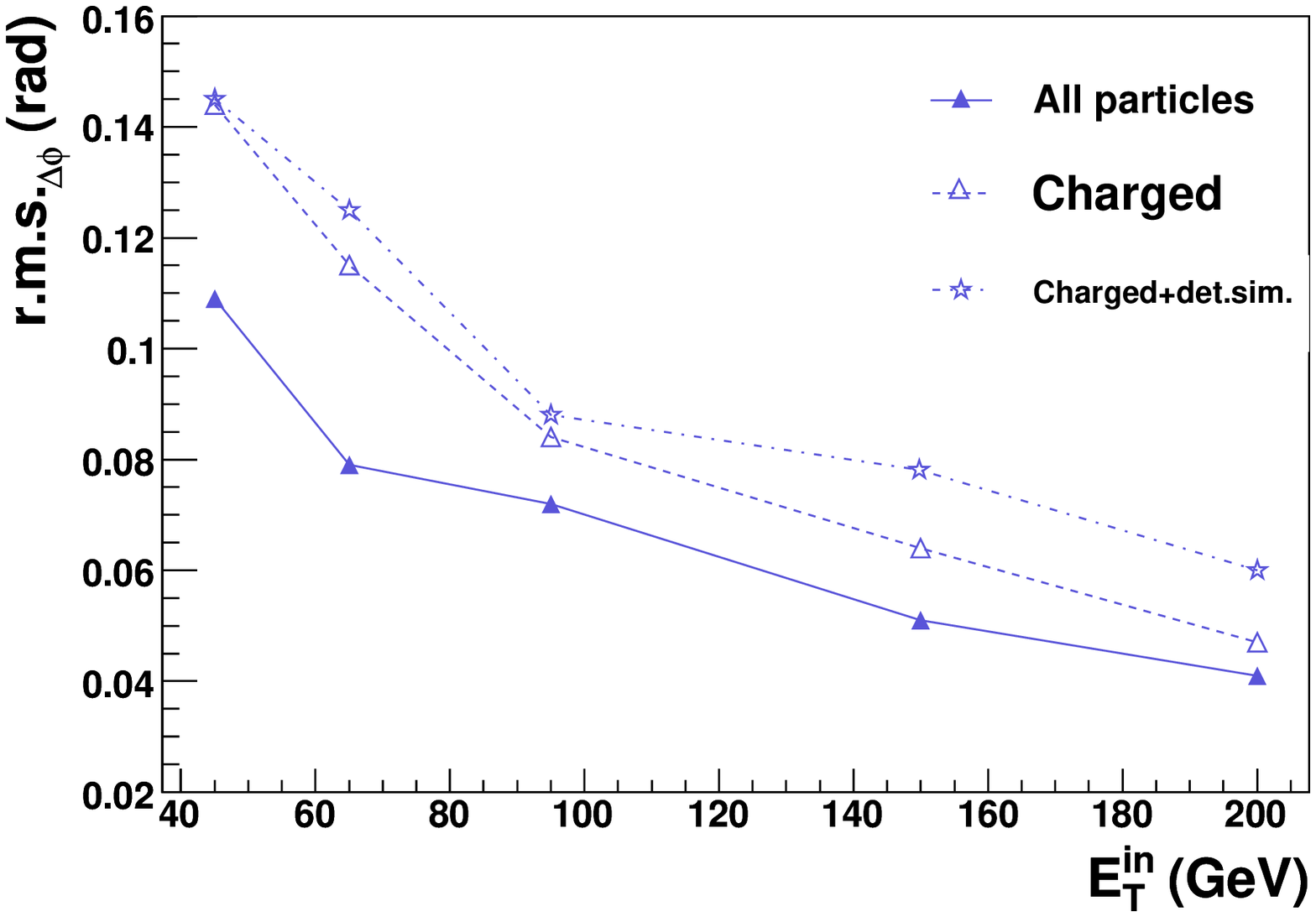}
   \caption{Left: Energy resolution of reconstructed jets as a function of different generated jet energy, when all
   particles (full line), charged particles (dashed line) are considered, and with fast detector response simulation
   (dotted line). Right: Direction resolution of reconstructed jets as a function of different generated jet energy.}
   \label{EtPhiRes}
\end{figure}

\begin{figure}
   \includegraphics[scale=0.45]{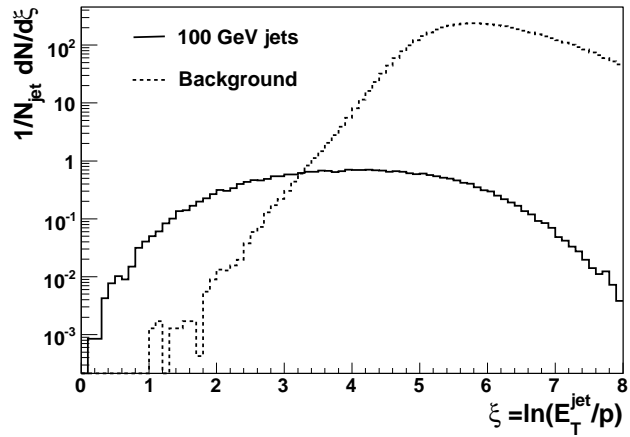}
   \caption{Fragmentation function for 100 GeV PYTHIA jets reconstructed with DA, compared with background
   coming from central Pb-Pb (HIJING) collisions.}
   \label{Fragm}
\end{figure}

\begin{figure}[th]
\vspace{0.5cm}
\includegraphics[width=0.6\textwidth]{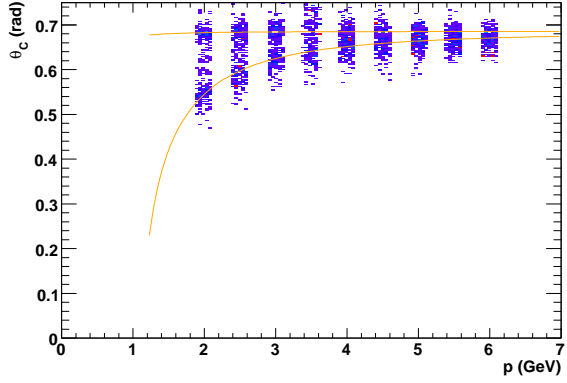}
\caption{Studied momentum regions are shown with the extracted
$\Theta_{C}$ for particle identification.} \label{figure:PreHbt1}
\end{figure}

\begin{figure}[th]
\vspace{0.5cm}
\includegraphics[width=0.45\textwidth]{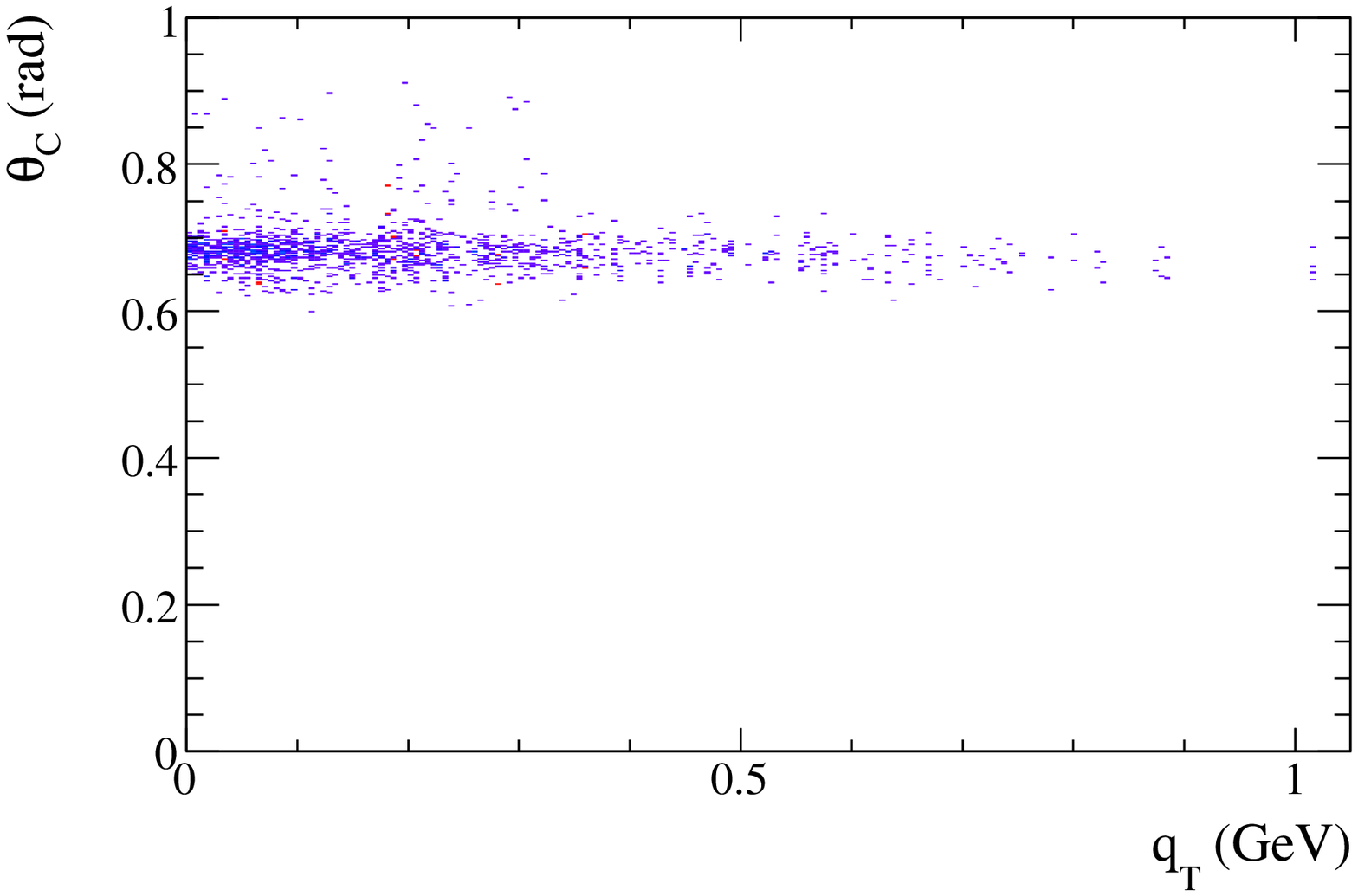}
\includegraphics[width=0.45\textwidth]{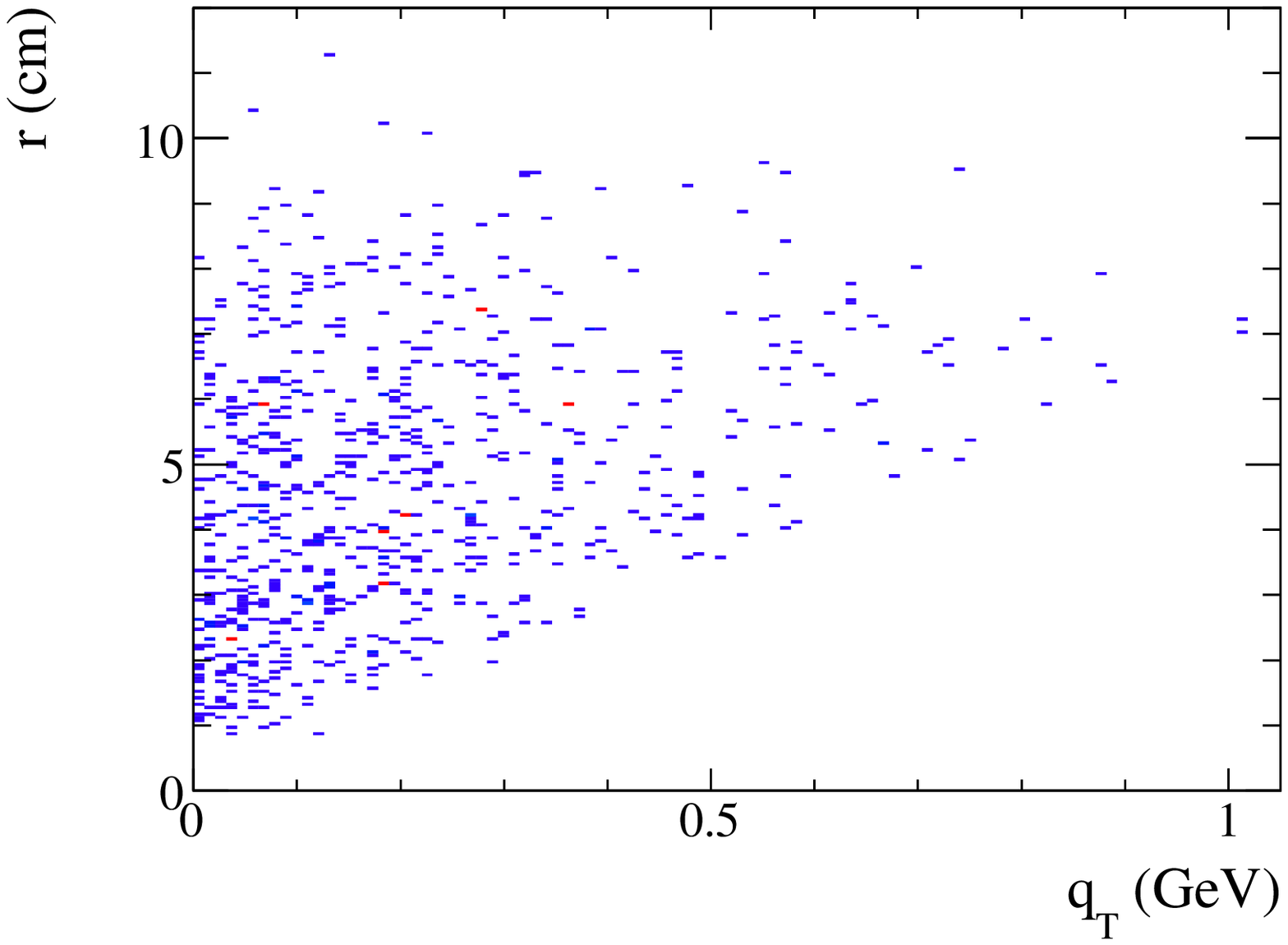}
\caption{$\Theta_{C}$ and distance of MIPS (tracks) as a function of
$q_T$ variables are shown.} \label{figure:PreHbt2}
\end{figure}

\section{Deterministic Annealing}

At LHC, the 98\% of total cross section will be represented by hard
processes. For this reason, jet production will be copious and jet
studies will be suitable to probe the hot and dense matter that will
be formed in heavy-ion collisions. A new algorithm, Deterministic
Annealing (DA), has been recently developed for jet finding in
ALICE. It satisfies important theoretical and experimental
requirements \cite{RunII} and it has been tested on simulated p-p
and Pb-Pb collisions. Results from the DA algorithm are shown in
Figure~\ref{EtPhiRes}, obtained from simulated PYTHIA jet events
embedded into Pb-Pb HIJING events, both generated at
$\sqrt{s_{NN}}=5.5$ TeV. The transverse energy and the direction
resolution of the reconstructed jets as a function of $E_{T}^{in}$
is referred to the same monochromatic input spectrum of jets,
reconstructed with the DA algorithm in a background free
environment. While for ideal response detectors analysis
(i.e.~considering all the particles) energy resolution always
improves with energy, charged-to-neutral fluctuations dominate when
ALICE reconstruction capabilities are considered. It can be noticed
that, even with fast simulation, energy resolution reaches at most
40\%, a value that is similar to the one achieved by cone algorithm
with a small radius \cite{cone}. As regards direction resolution, it
smoothly improves as the input jets transverse energy grows, as one
could expect.

HMPID will contribute to the study of jet properties. In fact, the
influence of medium on parton propagation after the hard scattering
can be studied comparing the leading particle fragmentation between
p-p and Pb-Pb collisions. In particular, from PYTHIA jet quenching
simulation an enhancement of this function in the low p$_T$ region,
and a corresponding depletion in the lower $\xi$ values region, more
visible for higher values of the transport coefficient $<\hat{q}>$,
can be observed \cite{AlicePPR}. In Figure~\ref{Fragm} the
fragmentation of jets found with Deterministic Annealing is compared
with background particles included in the same jet surface, coming
from Pb-Pb collisions. As a practical example, for 100 GeV jets
$\xi$ $\approx$ 3.5 corresponds to $p\approx$~3~GeV/$c$, that is
inside the HMPID $p_T$ coverage.

\section{HBT feasibility study}

This final-state information of heavy-ion collisions is more
directly accessed by interferometric methods, such as the
Hanbury-Brown and Twiss (HBT) measurement. At LHC, due to the high
multiplicity, event-by-event HBT could be performed. To achieve
event-by-event HBT measurements, 3$\sigma$ separation in particle
identification~\cite{AlicePPR} is required. The HMPID can contribute
in the high-p$_T$ region: to $\pi^{\pm}$$\pi^{\pm}$,
$K^{\pm}K^{\pm}$ and $p^{\pm}p^{\pm}$ HBT measurements. Furthermore,
non-identical correlations, such as $p$-$\Lambda$ could be performed
to investigate the contribution of flow effects. A preliminary study
on the HMPID detector resolution of two close tracks addresses the
capabilities of HMPID for HBT. Fast MC simulations of two tracks
($\pi$-$\pi$ and $p$-$p$) in close phase space (at least one pad
separation) and in various momentum bins are performed, including
complete simulation of HMPID response in a 0.2 T magnetic field. The
extracted $\Theta_{C}$ as a function of the studied momentum regions
is shown in Figure~\ref{figure:PreHbt1} for charged pions and
protons. Figure~\ref{figure:PreHbt2} shows the $\Theta_{C}$ for
charged pions and the extracted distance of two tracks for protons
as a function of $q_{T}$. Resolution down to $q_{T} \sim$ 10 $-$ 50
MeV is achievable with the default pattern recognition algorithm.

\section{Summary}
RHIC results have shown that QCD alone is not enough to describe the
full spectrum of hadronic physics in ultrarelativistic heavy-ion
collisions. Especially in the momentum range 2 - 5 GeV/$c$, where
both hard and soft processes come into play. Measurements in this
region could give insights to the processes that drive the collision
evolution. At LHC energies the High Momentum Particle Identification
detector will identify charged pions, kaon and protons and also
resonances such as $\phi$(1020) in the above momentum region.
Details of the HMPID design and its working principles were shortly
presented. Furthermore, selected physics topics of the HMPID were
discussed such as jet reconstruction via Deterministic Annealing and
particle correlations.

%


\begin{thebibliography}{00}

\bibitem{tdr} CERN / LHCC 98 - 19 ALICE TDR 1998.

\bibitem{AlicePPR}
  B.~Alessandro {\it et al.}  [ALICE Collaboration],
  J.\ Phys.\ G {\bf 32}, 1295 (2006).

\bibitem{coal}
  V.~Greco and C.~M.~Ko,
  Acta Phys.\ Hung.\  A {\bf 24}, 235 (2005)
  [arXiv:nucl-th/0405040] , \\
  B.~I.~Abelev {\it et al.}  [STAR Collaboration],
  Phys.\ Rev.\ Lett.\  {\bf 97}, 152301 (2006)
  [arXiv:nucl-ex/0606003].

\bibitem{phiProbe}
  A.~Shor,
  Phys.\ Rev.\ Lett.\  {\bf 54} (1985) 1122.

\bibitem{v2}
  B.~I.~Abelev {\it et al.} [STAR Collaboration],
  Phys.\ Rev.\ Lett.\  {\bf 99}, 112301 (2007)
  [arXiv:nucl-ex/0703033].

\bibitem{ppi}
  S.~S.~Adler {\it et al.}  [PHENIX Collaboration],
  Phys.\ Rev.\ Lett.\  {\bf 91}, 172301 (2003)
  [arXiv:nucl-ex/0305036].

\bibitem{pspec}
  K.~J.~Eskola, H.~Honkanen, H.~Niemi, P.~V.~Ruuskanen and S.~S.~Rasanen,
  Phys.\ Rev.\  C {\bf 72}, 044904 (2005)
  [arXiv:hep-ph/0506049].

\bibitem{RunII}
  G.~C.~Blazey {\it et al.},
  arXiv:hep-ex/0005012.

\bibitem{cone}
S.~L.~Blyth {\it et al.},
  J.\ Phys.\ G {\bf 34}, 271 (2007)
  [arXiv:nucl-ex/0609023].


\end{thebibliography}
\end{document}